\documentclass[twocolumn,prl,amsmath,amssymb,floatfix,superscriptaddress,showpacs]{revtex4}

\usepackage{times}
\usepackage{color}
\usepackage{graphicx}% Include figure files

\usepackage{dcolumn}% Align table columns on decimal point
\usepackage{bm}% bold math

\begin{document}

\title{Spin-wave induced phonon resonance in multiferroic BiFeO$_3$}

\author{Zhijun~Xu}
\affiliation{Condensed Matter Physics and Materials Science Department, Brookhaven National Laboratory,
	Upton, New York 11973, USA}
\affiliation{NIST Center for Neutron Research, National Institute of Standards and Technology, Gaithersburg, Maryland 20877, USA} 
\affiliation{Department of
	Materials Science and Engineering, University of Maryland, College Park, 
	Maryland, 20742, USA}
\affiliation{Physics Department, University of California, Berkeley,
	California 94720, USA} 
\affiliation{Materials Science Division,
	Lawrence Berkeley National Laboratory, Berkeley, California 94720,
	USA}

\author{J.~A.~Schneeloch}
\affiliation{Condensed Matter Physics and Materials Science
	Department, Brookhaven National Laboratory, Upton, New York 11973,
	USA} \affiliation{Department of Physics, Stony Brook University,
	Stony Brook, New York 11794, USA}

\author{Jinsheng~Wen}
\affiliation{Center for Superconducting Physics and Materials,
	National Laboratory of Solid State Microstructures and Department of
	Physics, Nanjing University, Nanjing 210093, China}
\affiliation{Physics Department, University of California, Berkeley,
	California 94720, USA} 
\affiliation{Materials Science Division,
	Lawrence Berkeley National Laboratory, Berkeley, California 94720,
	USA}

\author{M.~Matsuda}
\author{D.~Pajerowski}
\author{B. L. Winn}

\affiliation{Quantum Condensed Matter Division, Oak Ridge National
	Laboratory, Oak Ridge, Tennessee 37831, USA}

\author{Yang Zhao}
\affiliation{NIST Center for Neutron Research, National Institute of Standards and Technology, Gaithersburg, Maryland 20877, USA} 
\affiliation{Department of
	Materials Science and Engineering, University of Maryland, College Park, 
	Maryland, 20742, USA}

\author{Christopher Stock}
\affiliation{School of Physics and Astronomy, University of Edinburgh, Edinburgh EH9 3JZ, United Kingdom}

\author{Peter M. Gehring}
\affiliation{NIST Center for Neutron Research, National Institute of Standards and Technology, Gaithersburg, Maryland 20877, USA} 
\affiliation{Department of
	Materials Science and Engineering, University of Maryland, College Park, 
	Maryland, 20742, USA}

\author{T.~Ushiyama}
\author{Y.~Yanagisawa}
\author{Y.~Tomioka}
\author{T.~Ito}
\affiliation{National Institute of Advanced Industrial Science and Technology (AIST), Tsukuba, Ibaraki 305-8562, Japan}

\author{Genda~Gu}

\affiliation{Condensed Matter Physics and Materials Science
	Department, Brookhaven National Laboratory, Upton, New York 11973,
	USA}

\author{R. J. Birgeneau}
\affiliation{Physics Department, University of California, Berkeley,
	California 94720, USA} 
\affiliation{Materials Science Division,
	Lawrence Berkeley National Laboratory, Berkeley, California 94720,
	USA}

\author{Guangyong~Xu}
\affiliation{Condensed Matter Physics and Materials Science
	Department, Brookhaven National Laboratory, Upton, New York 11973,
	USA}
\affiliation{NIST Center for Neutron Research, National Institute of Standards and Technology, Gaithersburg, Maryland 20877, USA}



\begin{abstract}

We report the direct observation of a ``resonance'' mode in the lowest-energy optic phonon very near the zone center around (111) in the multiferroic BiFeO$_3$ using neutron scattering methods. The phonon scattering intensity is enhanced when antiferromagnetic (AFM) order sets in at T$_N = 640$~K, and it increases on cooling. This ``resonance'' is confined to a very narrow region in energy-momentum space where no spin-wave excitation intensity is expected, and it can be modified by an external magnetic field.  Our results suggest the existence of a novel coupling between the lattice and spin fluctuations in this multiferroic system in which the spin-wave excitations are mapped onto the lattice vibrations via the Dzyaloshinskii-Moriya (DM) interaction.
\end{abstract}

\pacs{63.20.kk,75.30.Ds,75.85.+t,75.25.-j,61.05.fg}
\maketitle

Having multiple ferroic orders coexisting in the same system, multiferroics provide fascinating platforms with which to study the interactions between different orders and, ultimately, to tailor these interactions in the pursuit of new technologies~\cite{datastorage,Eerenstein,swcheong}. In addition to coupling different order parameters, these interactions can also couple different excitations.  Phonons and magnons, which are quasiparticles arising from lattice vibrations and spin fluctuations, respectively, are two of the most fundamental excitations in solids.  When the coupling between these quasiparticles is sufficiently strong an overall modification of the phonon and/or magnon dispersion can be observed~\cite{Kim1988,Wagman2015,Oh2016}. This phenomenon is not limited to multiferroics, as it occurs in other materials as well.  Such interactions can even give rise to new types of hybrid quasiparticles, e.g. electromagnons~\cite{Pimenov,Cazayous,Kumar1,Rovillain}, which could be sensitive to more than one external parameter such as electric and magnetic fields, thereby offering tremendous potential for use in advanced device applications.

While quasiparticles are fundamentally quantum mechanical concepts, the interactions between them could lead to effects that occur in classical waves, such as the "resonance".
A resonance generally refers to a sharp increase of the amplitude of an oscillatory mode of a physical system at a specific frequency that is driven by interactions with another mode or an external force.  Interesting situations exist where the change of a physical quantity induces a resonance in a seemingly unrelated mode in the same system, thereby revealing a hidden correlation.  Well known examples of such induced resonances include the spin resonance~\cite{YBCO_resonance1,YBCO_Mook,Christianson2008,Stock115} in unconventional superconductors and the phonon resonance~\cite{Allen1997,Weber} in conventional superconductors, which suggest that spin/lattice fluctuations play important roles in the superconducting pairing mechanism. 

\begin{figure}
	\includegraphics[width=0.9\linewidth]{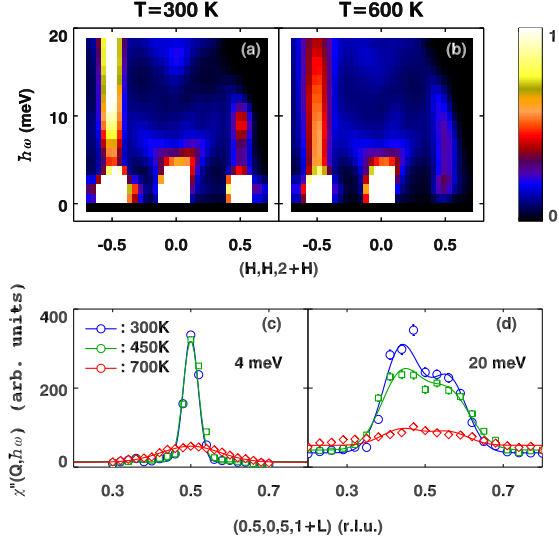}
	\caption{Magnon and phonon neutron scattering cross sections in BFO.  Data measured on HYSPEC at (a) 300~K and (b) 600~K are plotted in the energy-momentum plane near the nuclear Bragg peak (002) for {\bf Q} along [111].  Relative intensities are shown on a linear color scale given by the vertical bar at right.  Linear intensity profiles measured on HB1 at (c) 4~meV and (d) 20~meV are shown as a function of {\bf Q} scanned along [001] and centered on the magnetic Bragg peak (0.5,0.5,1.5).  All intensities have been divided by the Bose factor to obtain $\chi"$. Error bars represent $\pm 1$ standard deviation. }\label{fig:1}
\end{figure}

BiFeO$_3$ (BFO) is an extensively studied multiferroic material with superior dielectric properties in which antiferromagnetic and ferroelectric order coexist at room temperature~\cite{Catalan}.  The coupling between static magnetic and ferroelectric order in BFO results in a cycloid spin structure~\cite{spiral} with a large modulation ($\sim 620~\AA$), and small anomalies in the phonon energies at the magnetic phase transition have been reported as well~\cite{haumont:132101,Rovillain_Phonon}.  Both theoretical~\cite{sousa:012406} and experimental (Raman) studies~\cite{Cazayous,Kumar1,Rovillain,BiFeO3_film_field} suggest the existence of hybrid quasiparticles, but no evidence of these has been found by neutron scattering~\cite{Delaire,cheong,Matsuda,BFO_Xu,Schneeloch2015}.  We report neutron scattering data that indicate that the energies and intensities of the low-energy phonons in BFO do not appear to be affected by the magnetic order throughout most of reciprocal space, in good agreement with previous work~\cite{Schneeloch2015}.  However, our data also reveal a striking exception to this scenario in that the scattering intensity from the lowest-energy optic phonon very near the zone center ${\bf Q}=(1,1,1)$ at $\hbar\omega \sim 7$~meV increases sharply below the N{\'e}el temperature T$_N$.  This phonon resonance can be explained using a simple phenomenological model that takes into account the DM interactions between spin fluctuations, effectively mapping the spin-wave excitations onto the lattice vibrations.  This yields a novel dynamic coupling mechanism in which (i) the intensity of one quasiparticle is greatly enhanced, but only in a small, well-defined region of energy-momentum space, while all other quasiparticle properties (i.\ e.\ dispersions, intensities) are unaffected; (ii) the resonance occurs at an energy-momentum position where one quasiparticle has strong spectral weight while the other does not; and (iii) the resonance intensity adds to the original quasiparticle intensity instead of appearing as a new mode.  Our data thus reveal a new and unique example of how phonons and magnons can interact in multiferroic systems.  Measurements made in an external magnetic field provide further confirmation of this unusual coupling, as the resonance intensity increases with field as does the ferroelectric polarization along [111].  This behavior suggests that the same type of DM interaction that couples the static spin and polar structures is also behind the induction of this highly coherent resonance mode.

The low-energy magnon and phonon excitations in BFO can be measured simultaneously with neutron inelastic scattering~\cite{Note_BFO} on single crystal samples~\cite{Ito_BFO,Kawachi2017}.  In Fig.~1 we plot the imaginary part of the generalized susceptibility $\chi''({\bf Q},\hbar\omega)$ obtained from measurements made near the (002) nuclear Bragg peak and for wave vectors along [111] at (a) 300~K and (b) 600~K.  This is obtained by dividing the raw scattering intensity $I({\bf Q},\hbar\omega)$ by the Bose factor $\frac{1}{1-e^{-\hbar\omega/k_BT}}$.  The magnon contribution to $\chi''({\bf Q},\hbar\omega)$ decreases and broadens in energy on heating above T$_N$ (640~K) as expected~\cite{BFO_Xu}.  This is evident from panels (c) and (b), which show data measured near the AFM wave vector (-0.5,-0.5,1.5) at energies of 4~meV and 20~meV, respectively, and which is consistent with the fast decay of magnons once the static magnetic order dissolves.  For temperatures well below T$_N$ the magnetic scattering intensity follows the Bose factor.  The lattice contribution to $\chi''({\bf Q},\hbar\omega)$ in panels (a) and (b) exhibits no significant changes between 300~K and 600~K, thus indicating that the phonon scattering also follows the Bose factor.  This behavior is also expected and holds almost everywhere in reciprocal space that we probed. A more quantitative and extensive survey is provided in Fig.~2 where the scattering from longitudinal and transverse phonons in different Brillouin zones and selected wave vectors are plotted.  With one exception (Fig.~2(a)), the phonon spectra exhibit a normal temperature dependence where energies slightly harden on cooling into the ferroelectric phase and intensities follow the Bose factor.  However, when we examined the zone-center ($q=0$) optic mode at (111), we discovered a significant and unexpected intensity enhancement, or resonance, for $T < T_N$.

\begin{figure}
	\includegraphics[clip,trim=0cm 1cm 0cm 0cm,width=0.9\linewidth]{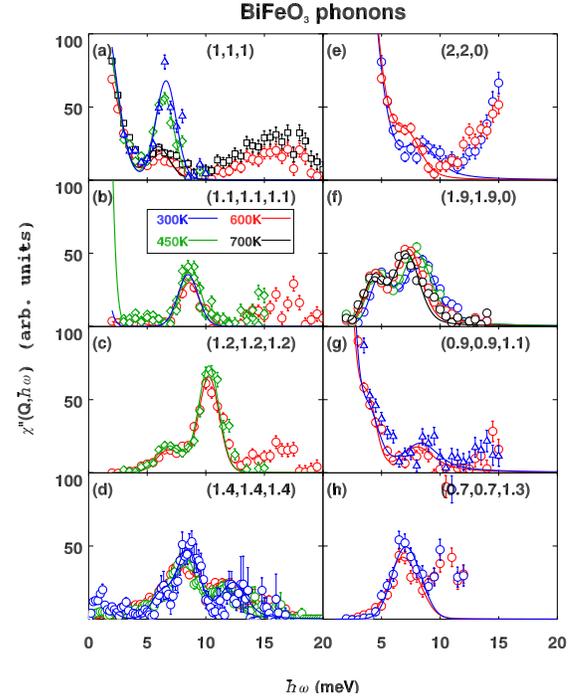}
	\caption{$\chi''({\bf Q},\hbar\omega)$ derived from energy scans performed at constant {\bf Q}. (a)-(d) Longitudinal phonons measured near (111). (e) Zone-center phonon measured at (220). (Red data points here correspond to 650~K instead of 600~K.)  (f) Transverse phonon (q=0.1) measured near (220). (g) and (h) Transverse phonons measured near (111). Lines are guides to the eye.  All data were taken on HB-1 except for those in (d) which were measured on HYSPEC with a $Q$-integration width of 0.1 (r.l.u.)}\label{fig:2}
\end{figure}

To better illustrate this resonance response, we plot the ratio of the energy-integrated phonon intensity, divided by the Bose factor, to that of the same mode measured at 600~K in Fig.~3 for a number of selected reciprocal space locations.  This ratio rises quickly on cooling below T$_N$ at ${\bf Q}=(1,1,1)$ (zone center optic mode) and nearly triples in size by 300~K.  But if we displace the wave vector {\bf Q} away from (1,1,1) by a small amount along either the longitudinal or a transverse direction, the resonant response disappears.  The ratio measured at other locations always remains close to one for both optic and acoustic modes.  Surprisingly, the resonance is only observed clearly at ${\bf Q}=(1,1,1)$.  It is much reduced, if not completely suppressed, at the equivalent positions ${\bf Q}=(2,2,0)$ and $(0,0,2)$. Since our neutron scattering measurements are sensitive to phonons polarized along ${\bf Q}$, this implies that the intensity enhancement is associated primarily with atomic displacements (polarizations) along $\langle111\rangle$.

\begin{figure}
	\includegraphics[width=0.9\linewidth]{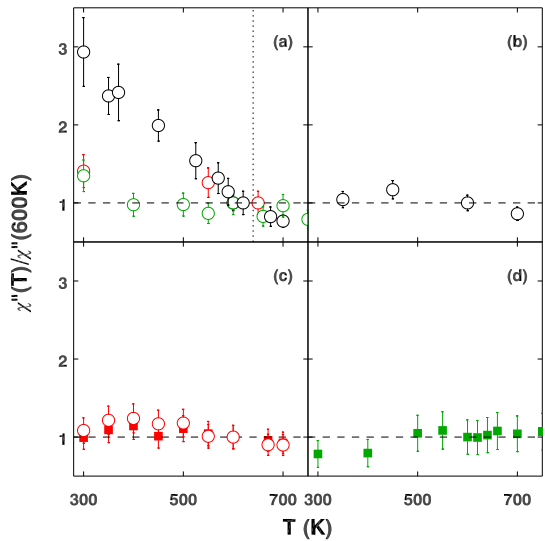}
	\caption{Temperature dependence of $\chi"(T)/\chi"(600K)$ for various phonons. (a) Zone-center ($q=0$) optic phonons at (111) (black), (220) (red), and (200) (green). (For (220) the ratio $\chi"(T)/\chi"(650K)$ is plotted because no 600~K data were measured.)  The vertical dotted line denotes T$_N=640$~K. (b) LO phonon at {\bf Q}=(1.1,1.1,1.1). (c) LA (red squares) and LO (red circles) phonons at {\bf Q}=(1.9,1.9,0). (d) TA phonon at {\bf Q}=(2,0.4,0). Error bars represent fitting errors.}\label{fig:3}
	
\end{figure}

There are several possible origins of the resonance.  One is the coupling that exists between the magnetic order and the phonon modes, i.\ e.\ magneto-elastic coupling that gives rise to ``magneto-vibrational" modes~\cite{Aksenov1981,Brown1990}.  In this case both phonon dispersions and intensities can be modified by the onset of magnetic order.  However, this type of coupling generally affects the phonon energies and intensities over a large range of momentum, not at a single value of ${\bf Q}$ and energy as is observed in our case.  Another possibility is a direct coupling between a spin-wave excitation and an optic phonon mode where they cross each other~\cite{Wagman2015}.  And in fact, the spin-wave dispersion in BFO~\cite{BFO_Xu} does approach the low-energy optic phonon branch near $q=0$ and $\hbar\omega \sim 7$~meV.  However, the spectral weight of the spin-wave excitations in a G-type AFM system is governed by a structure factor that is typically largest at the AFM wave vector $q=\pi$ (the AFM zone center) and zero at $q=0$ (the nuclear zone center).  While in principle such a direct coupling between the two modes at this crossing point cannot be ruled out, it is difficult to account for the large spectral enhancement of the original optic phonon mode near $q=0$ when the crossing spin wave has almost zero spectral weight.

Instead, we propose a scenario in which the coupling between two static order parameters (ferroelectric and AFM) can be extended to couple excitations arising from fluctuations of these order parameters.  The static spin structure in BFO (see Fig.~4a) consists of spins ordered antiferromagnetically along [111] with $\vec{Q}_{AF}=(0.5,0.5,0.5)$ ($q=\pi$) and a cycloid order with long modulation along the perpendicular $\langle01\bar{1}\rangle$ directions~\cite{spiral}.  When a spontaneous ferroelectric polarization is present, the free energy of the system can be lowered if the polarization ${\vec P_0}$ is parallel to $\vec{L}\times (\vec{S_i} \times \vec{S}_{i+1})$ because of the antisymmetric exchange (DM interaction) between neighboring spins~\cite{Yoshinori2014}. Here $\vec{L}$ is the vector connecting two adjacent spins $\vec{S_i}$ and $\vec{S}_{i+1}$ along the direction of the cycloid.  This effectively couples the ferroelectric polar order at $q=0$ to the AFM cycloid spin structure at $q=\pi$ along [111]. The reason behind this mapping of nuclear zone center ($q=0$) to AFM zone center ($q=\pi$), which is also the nuclear zone boundary, is that the DM interaction is only sensitive to the chirality of the cycloid. The two anti-phase rows of spins shown in Fig.~4a have the same chirality; therefore the DM interaction couples the AFM spin structure, with a periodicity of two sites along [111], to the spontaneous polarization, which has half the periodicity, i.\ e.\ just one site along [111].

When a spin-wave excitation forms near $q=\pi$ from fluctuations of the ground state spin configuration, one would naturally expect the lattice polarization to fluctuate in a related manner due to the DM coupling, leading to a lattice mode near $q=0$.  In the Supplemental Materials we give a simplified picture of how a ``magneto-phonon" mode can be induced by such a dynamic DM interaction, having a linear dispersion near $q=0$ similar to that of the spin wave near $q=\pi$ (see Fig.~4b). More detailed theoretical work is definitely required to develop a quantitative understanding of this novel dynamic coupling. It is also important to understand that this is a mapping of one magnon near $q=\pi$ to a lattice mode around $q=0$, which differs from the creation of a two-magnon mode near $q=0$ by combining two magnons near $q=\pi$.

\begin{figure}
	\includegraphics[width=0.9\linewidth]{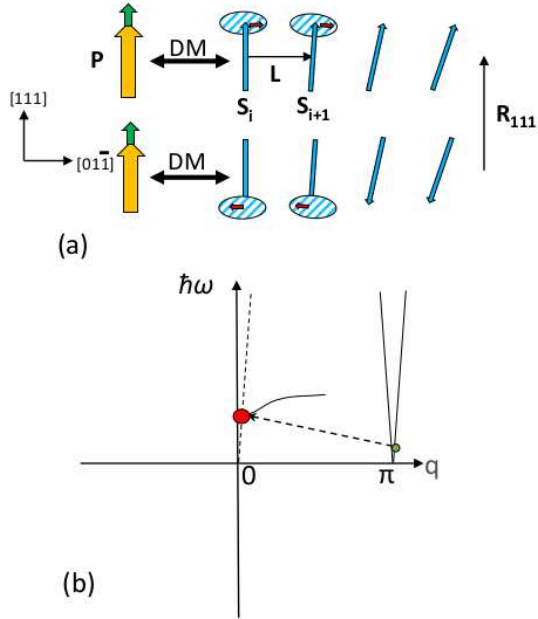}
	\caption{Schematic diagram showing how spin-wave excitations near $q=\pi$ can be mapped onto lattice vibrational modes near $q=0$. (a) Spins in the cycloid plane and the corresponding lattice polarization along [111]. The small arrows are fluctuating (precessing) spins (red) and fluctuating polarizations (green). (b) A spin wave mode at $(\pi+q,\hbar\omega)$ is mapped to a lattice mode at $(2q,2\hbar\omega)$. Solid lines represent the spin-wave (near $q=\pi$) and optic phonon (near $q=0$) dispersions, and the dashed line represents the magneto-phonon mode mapped from the spin-wave dispersion.}\label{fig:4}
\end{figure}

The strength of the induced ``magneto-phonon" mode depends on many factors including the magnitude of the spin fluctuation $\Delta S$, the strength of the antisymmetric interaction, and the presence of a non-collinear spin structure (i.\ e.\ $\vec{S}_i\times\vec{S}_{i+1}\neq 0$).  It should vanish upon warming above T$_N$ when the N{\'e}el order melts.  It is a second order effect that is not expected to be strong, which explains why previous neutron and x-ray scattering studies did not detect this mode near $q=0$.  However, a resonance can occur if another mode crosses the induced mode so that at the crossing point the two modes have the same energy and momentum.  If we examine Fig.~4b, we can see that this magneto-phonon mode crosses the lowest-energy polar optic phonon branch in BFO around $7~$meV for $q\sim 0.02~\AA^{-1}$ to $0.05~\AA^{-1}$ (depending on the direction of $\vec{q}$) based on the known spin-wave dispersion~\cite{BFO_Xu}.  This naturally explains why optic probes such as Raman spectroscopy~\cite{BFO_Raman1,BFO_Raman2,haumont:132101,Rovillain_Phonon} or infrared spectroscopy~\cite{Lobo_infrared}, which are sensitive only to modes at $q=0$, are unable to detect the resonance: the crossing occurs at a small but non-zero $q$ as discussed above.  The reason why the resonance appears at $q=0$ in our neutron inelastic scattering measurements is that the instrumental wave vector resolution is comparatively coarse near (111) ($\sim 0.1~\AA^{-1}$ in the transverse direction).  Thus a nominal scan at $q=0$ will pick up the resonance intensity because it is located at a wave vector that lies within the instrumental resolution.  Another consequence is that the resonance is only pronounced near (111) since the spin excitations are only coupled to $\langle111\rangle$ polarization fluctuations, and the resonance will likely only occur for optic phonons having the same polarizations as the magneto-phonon mode.

Note that the exact form of the interaction is not essential here - the central requirement is that the same type of coupling (which is assumed to be an asymmetric DM interaction) between the static spin and polar structures is also present between the spin and lattice fluctuations; this results in a significant enhancement, or resonance, of the polar phonon intensity at a specific wave vector and energy.  Evidence supporting the validity of this model is observed when a large external magnetic field is applied along [1$\bar{1}$0], which is known to induce a magnetic phase transition~\cite{Tokunaga2015,Kawachi2017} (see the field-dependent changes in the magnetic Bragg peak shown in Fig.~S1a).  While the exact spin structure of the high field phase of BFO is unknown, which is unlikely a simple collinear AFM phase, the ferroelectric polarization along [111] increases significantly with field~\cite{Tokunaga2015}, suggesting an enhancement of the asymmetric spin interaction.  Consistent with the picture described above, we find that the zone center ($q=0$) phonon is also enhanced by the field whereas the $q=0.1$ phonon remains the same in all phases (see Fig.~5b and 5c). 

\begin{figure}
	\includegraphics[clip,trim=0cm -2cm 1cm -1.5cm,width=0.9\linewidth]{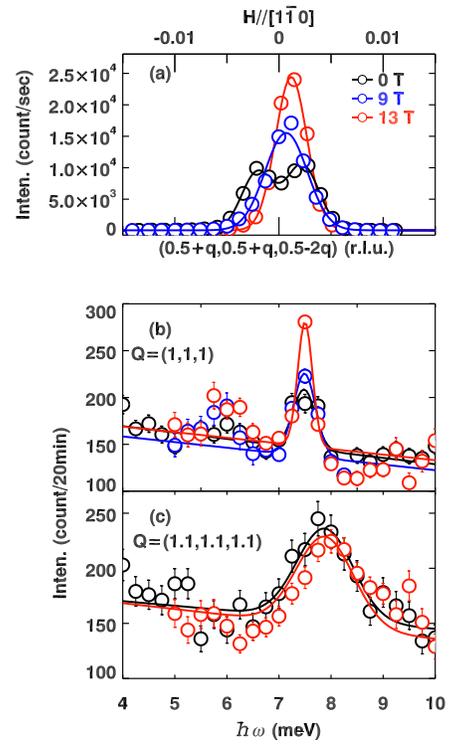}
	\caption{Magnetic Bragg peak and (111) phonon measured under external magnetic field at 300~K on BT7. (a) Magnetic Bragg peak (0.5,0.5,0.5). (b) and (c) Longitudinal phonons measured near (111) for q=0 and q=0.1, respectively.}\label{fig:5}
	
\end{figure}

We have discovered a magnetically induced enhancement of the lowest-energy zone-center optic phonon at (111) in BFO.  This enhancement is located far from any strong magnetic excitation or magnetic Bragg peak, which indicates that this effect is due to an entirely new mechanism that couples lattice and spin fluctuations.  Our results suggest that, in addition to the commonly observed coupling between static spins and ferroelectric polarizations, the asymmetric spin interactions in this multiferroic system contribute to a highly unusual resonance that occurs only at reciprocal space locations where a polar phonon mode coincides with a ``magneto-phonon mode" that is induced by the spin excitations, and they do so without affecting the phonon or magnon dispersions.

\begin{acknowledgments} ZJX, JAS, GDG, and GYX acknowledge support by Office of Basic Energy Sciences, U.S. Department of Energy under contract No. DE-SC0012704. JW and RJB are also supported by Work at Lawrence BerkeleyLaboratory and UC Berkeley was supported by the
	Office of Basic Energy Sciences (BES), Materials Sciences and
	Engineering Division of the U.S. Department of Energy (DOE) under
	Contract No. DE-AC02-05-CH1231 within the Quantum Materials Program
	(KC2202) and BES. A portion of this research used resources at the High Flux Isotope Reactor and the Spallation Neutron Source, DOE Office of Science User Facilities operated by the Oak Ridge National Laboratory. CS acknowledges the Carnegie Trust for the Universities of Scotland and the Royal Society.  TI is partly supported by the Mitsubishi Foundation.
\end{acknowledgments}

%\bibliography{../../References/BFO.bib,../../References/fetese.bib}

\end{document}